\documentclass[english,final,superscriptaddress]{revtex4}
\usepackage[T1]{fontenc}
\usepackage[latin9]{inputenc}
\usepackage{graphicx}
\usepackage{amssymb}

\makeatletter

\providecommand{\tabularnewline}{\\}

\@ifundefined{definecolor}
 {\usepackage{color}}{}

\makeatother

\usepackage{babel}

\begin{document}

\title{INDIRECT SEARCH FOR COLOR OCTET ELECTRON AT NEXT GENERATION LINEAR
COLLIDERS}

\author{A.N. AKAY}

\email{aakay@etu.edu.tr}

\address{TOBB University of Economics and Technology, Physics Division, Ankara,
Turkey}

\author{H. KARADENIZ}

\email{hande.karadeniz@taek.gov.tr}

\address{Turkish Atomic Energy Authority, SANAEM, Ankara, Turkey}

\author{M. SAHIN}

\email{m.sahin@etu.edu.tr}

\address{TOBB University of Economics and Technology, Physics Division, Ankara,Turkey}

\author{S. SULTANSOY}

\email{ssultansoy@etu.edu.tr}

\address{TOBB University of Economics and Technology, Physics Division, Ankara,
Turkey}

\address{National Academy of Sciences, Institute of Physics, Baku, Azerbaijan}
\begin{abstract}
In this study we investigated indirect manifestations of color octet
electron at the next generation linear colliders: International Linear
Collider (ILC) and Compact Linear Collider (CLIC). Namely, production
of two gluons via color octet electron exchange is considered. Signal
and background analysis have been performed taking into account initial
state radiation and beamstrahlung. We show that color octet electron
($e_{8}$) manifestation will be seen upto $M_{e_{8}}=1.75$ TeV and
$1.70$ TeV at ILC and CLIC with $\sqrt{s}=0.5$ TeV, respectively.
CLIC with {\normalsize $\sqrt{s}=3$ TeV will be sensitive upto }$M_{e_{8}}=6.88${\normalsize {}
TeV.} 
\end{abstract}
\maketitle

\section*{1. INTRODUCTION}

Predictions of Standard Model (SM) have so far been in agreement with
results of numerous experiments. Therefore, SM has been in use to
explain many puzzles related to fundamental particles and their interactions.
However, there have been a number of fundamental problems (quark-lepton
symmetry, family replication, hierarchy problems, charge quantization
etc.) SM could not deal with. Furthermore, masses and mixings of leptons
and quarks are fixed by hand. These problems forced physicists to
go beyond SM. Extra dimensions, supersymmetry (SUSY), compositeness
and so on have been created for solving these problems and to explain
the physics underlying them.

Among these alternative models compositeness has explained the subjects
of family replication and quark-lepton symmetry in the best manner
According to the theory of compositeness quarks, leptons, gauge bosons
need to be composite particles made up of more basic constituents.
These basic constituents, in otherwords substructural particles, are
named preons. It is assumed that preons are what compose quarks, leptons
and gauge bosons. It's also supposed that preons interactions lead
to development of many new types of particles such as leptoquarks,
leptogluons, diquarks, dileptons and excited fermions.

We will be interested in color octet leptons that in the framework
of composite models with colored preons \cite{I.A. D'SOUZA,HModel,FM Model,GS Model,BM Model,BS Model,CKS Model}
have the same status as the excited leptons. For instance, in the
framework of fermion-scalar models, leptons would be a bound state
of one fermionic preon and one scalar anti-preon $l=(F\bar{S})=1\oplus8$
(both $F$ and $S$ are color triplets), then each SM lepton is thought
to accompany with its own color octet partner \cite{CKS Model}. There
are many papers about manifestations of excited leptons at high energy
colliders whereas color octet leptons are discussed only in a few
papers \cite{Hewett,CKS Model,Kantar 1998,Sahin2010}.

Resonant production of $e_{8}$ at future ep colliders have been analyzed
in recent paper \cite{Sahin2010}, where also current limits on leptogluon
masses are briefly discussed. It was shown that the $e_{8}$ discovery
at the LHeC simultaneously will determine the compositeness scale.

In this paper we will consider indirect manifestation of color octet
electrons at the next generation linear colliders: International Linear
Collider (ILC) and Compact Linear Collider (CLIC). In section 2, we
present the interaction Lagrangian of the leptogluons as well as signal
cross-section for the process $e^{-}e^{+}\rightarrow gg$ via t-channel
$e_{8}$ exchange. The signal and background analysis performed at
ILC and CLIC is given in section 3. In the last section we give an
interpretation of obtained results and concluding remarks.

\section*{2. INTERACTION LAGRANGIAN AND INDIRECT PRODUCTION CROSS SECTIONS}

The interaction Lagrangian of leptogluons with the corresponding lepton
and gluon is given by \cite{PDG2010,Kantar 1998,Sahin2010}:

\begin{equation}
L=\frac{1}{2\Lambda}{\displaystyle \sum_{l}\{\bar{l_{8}}g_{s}G_{\mu\nu}^{\alpha}\sigma^{\mu\nu}(\eta_{L}l_{L}+\eta_{R}l_{R})+h.c.\}}\end{equation}

where $G_{\mu\nu}^{\alpha}$ is field strength tensor for gluon, index
$\alpha=1,2,...,8$ denotes the color, $g_{s}$ is gauge coupling,
$\eta_{L}$ and $\eta_{R}$ are the chirality factors, $l_{L}$ and
$l_{R}$ denote left and right spinor components of lepton, $\sigma^{\mu\nu}$
is the anti-symmetric tensor and $\Lambda$ is the compositeness scale,
which is taken to be equal to $M_{e_{8}}$. The leptonic chiral invariance
implies $\eta{}_{L}$$\eta_{R}=0$. For numerical calculations we
implement this Lagrangian into the CalcHEP program \cite{CalcHEP}.

The cross sections for the process $e^{-}e^{+}\rightarrow gg$ via
t-channel $e_{8}$ exchange at $\sqrt{s}=0.5$ TeV are presented in
Figure 1. ISR and Beamstrahlung effects at ILC and CLIC are calculated
with CalcHEP program using parameters given in Table I \cite{ILC,CLIC,CLICWEB}.

\begin{table}
\begin{tabular}{|c|c|c|c|}
\hline 
Collider Parameters  & ILC  & CLIC1  & CLIC2\tabularnewline
\hline
\hline 
$E(\sqrt{s})$ TeV  & $0.5$  & $0.5$  & $3$\tabularnewline
\hline 
$L(10^{34}cm^{-2}s^{-1})$  & $2$  & $2.3$  & $5.9$\tabularnewline
\hline 
$N$$(10^{10})$  & $2$  & $0.68$  & $0.372$\tabularnewline
\hline 
$\sigma_{x}$ (nm)  & $640$  & $202$  & $45$\tabularnewline
\hline 
$\sigma_{y}$ (nm)  & $5.7$  & $2.3$  & $1$\tabularnewline
\hline 
$\sigma_{z}$ ($\mu$m)  & $300$  & $44$  & $44$\tabularnewline
\hline
\end{tabular}

\caption{Main parameters of ILC and CLIC. Here N is the number of particles
in bunch. $\sigma_{x}$ and $\sigma_{y}$ are RMS beam sizes at Interaction
Point (IP), $\sigma_{z}$ is the RMS bunch length. }

\end{table}

In Figure 2 we present the results of similar calculations for the
CLIC2 $\sqrt{s}=3$ TeV. It is seen that ISR and Beamstrahlung effects
essentially reduce the cross sections.

\begin{figure}
\includegraphics[scale=0.7]{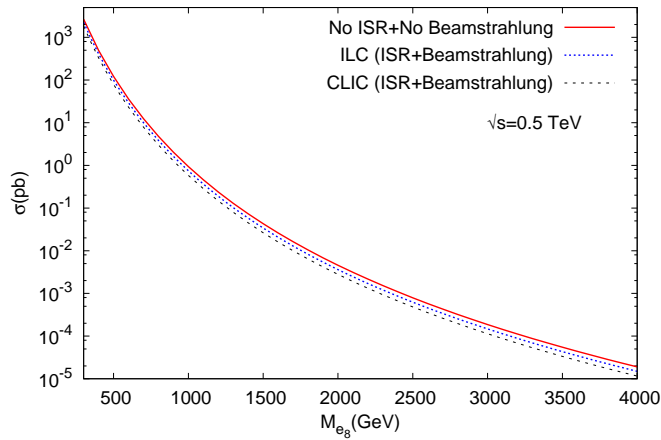}

\caption{Indirect production cross sections for color octet electron at the
ILC and CLIC1.}

\end{figure}

\begin{figure}
\includegraphics[scale=0.7]{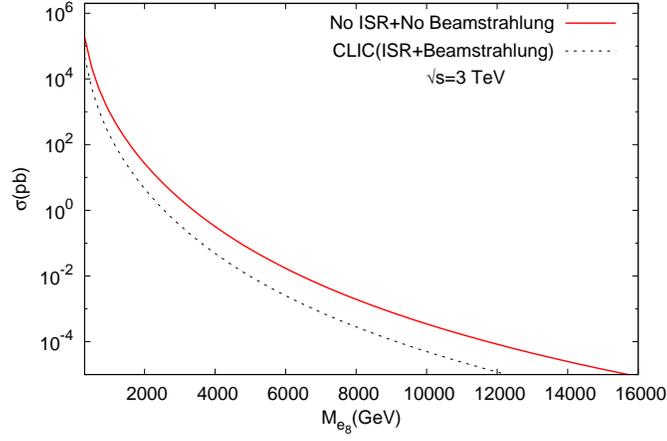}

\caption{Indirect production cross sections for color octet electron at the
CLIC2.}

\end{figure}

Figures 1 and 2 show that indirect manifestation of $e_{8}$ could
be seen for $e_{8}$ mass values several times higher than those for
direct pair production at corresponding center of mass energies, namely,
$0.25$ TeV at ILC/CLIC1 and $1.5$ TeV at CLIC2.

\section*{3. SIGNAL AND BACKGROUND ANALYSIS}

Our signal process is $e^{-}e^{+}\rightarrow gg$ and background processes
are $e^{-}e^{+}\rightarrow\gamma,Z\rightarrow jj$, where $j=u,\bar{u},d,\bar{d},c,\bar{c},s,\bar{s},b,\bar{b}$.
In order to determine appropriate cuts we calculate $p_{T}$, $\eta$
and invariant mass distributions for signal and background processes.
Results for ILC are presented in figures 3, 4 and 5, respectively.
It is seen that selection cuts $p_{T}>120$ GeV, $|\eta|<1$ and $375$
GeV $<M_{inv}(jj)<500$ GeV essentially suppress the background, whereas
the signal remains almost unchanged.

\begin{figure}
\includegraphics[scale=0.7]{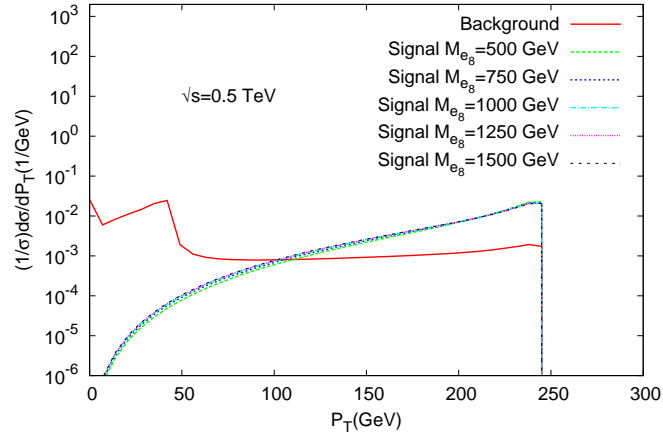}

\caption{Normalized transverse momentum distributions of final state jets for
signal and background at the ILC.}

\end{figure}

\begin{figure}
\includegraphics[scale=0.7]{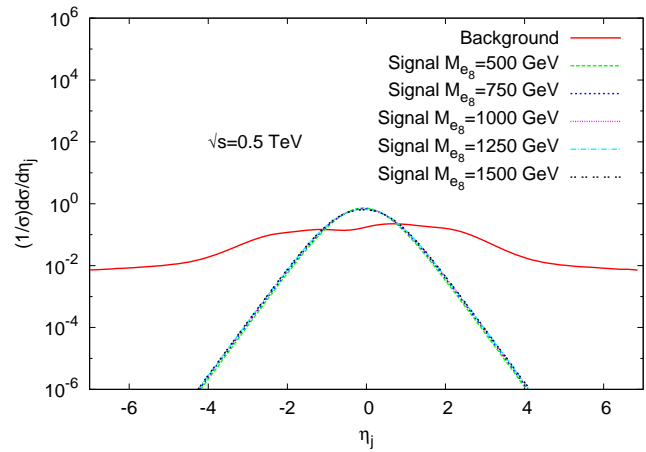}

\caption{Normalized pseudo-rapidity distributions of final state jets for signal
and background at the ILC. }

\end{figure}

\begin{figure}
\includegraphics[scale=0.7]{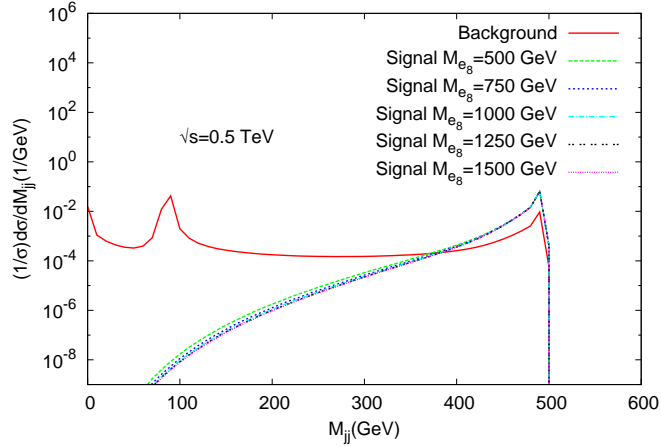}

\caption{Normalized invariant mass distributions of final state jets for signal
and background at the ILC.}

\end{figure}

As a result of similar analysis for CLIC1 and CLIC2 we will use following
sets of selection cuts: $p_{T}>120$ GeV, $|\eta|<1$ and $350$ GeV
$<M_{inv}(jj)<500$ GeV for CLIC1; $p_{T}>350$ GeV, $|\eta|<1$ and
$1000$ GeV $<M_{inv}(jj)<3000$ GeV for CLIC2.

In the Table II, we present signal and background cross-sections for
the ILC (CLIC1) after above-mentioned cut sets. Corresponding values
for CLIC2 are given in the Table III.

\begin{table}
\begin{tabular}{|c|c|c|}
\hline 
$M_{e_{8}}$, TeV  & \multicolumn{2}{c||}{$\sigma$, pb}\tabularnewline
\hline
\hline 
 & ILC  & CLIC1\tabularnewline
\hline 
$0.5$  & $89.835$  & $70.326$\tabularnewline
\hline 
$1$  & $0.663$  & $0.5091$\tabularnewline
\hline 
$1.5$  & $0.030$  & $0.0229$\tabularnewline
\hline 
$2$  & $0.0032$  & $0.0024$\tabularnewline
\hline 
Background  & $1.757$  & $1.763$\tabularnewline
\hline
\end{tabular}

\caption{Signal and background cross-sections after cuts for ILC and CLIC1.
These cuts are $p_{T}>120$ GeV, $|\eta|<1$ and $375$ GeV $<M_{inv}(jj)<500$
GeV for ILC and $p_{T}>120$ GeV, $|\eta|<1$ and $350$ GeV $<M_{inv}(jj)<500$
GeV for CLIC1. }

\end{table}

\begin{table}
\begin{tabular}{|c|c|}
\hline 
$M_{e_{8}}$, TeV  & $\sigma$, pb\tabularnewline
\hline
\hline 
$2$  & $4.0081$\tabularnewline
\hline 
$3$  & $0.3078$\tabularnewline
\hline 
$4$  & $0.0424$\tabularnewline
\hline 
$5$  & $0.0084$\tabularnewline
\hline 
$6$  & $0.00217$\tabularnewline
\hline 
$7$  & $0.00067$\tabularnewline
\hline 
$8$  & $0.00024$\tabularnewline
\hline 
Background  & $0.0378$\tabularnewline
\hline
\end{tabular}

\caption{Signal and background cross-sections after cuts for CLIC2. These cuts
are $p_{T}>350$ GeV, $|\eta|<1$ and $1000$ GeV $<M_{inv}(jj)<3000$
GeV. }

\end{table}

For statistical significance, we use following formula:

\begin{equation}
S=\frac{\sigma_{s}}{\sqrt{\sigma_{s}+\sigma_{b}}}\sqrt{L_{int}}\end{equation}

where $\sigma_{s}$ is signal cross-sections, $\sigma_{b}$ is background
cross-sections and $L_{int}$ is integrated luminosity. In Figures
6, 7 and 8 the necessary integrated luminosities for $2\sigma$ exclusion,
$3\sigma$ observation and $5\sigma$ (indirect) discovery of $e_{8}$
are plotted as a function of $e_{8}$ mass for ILC, CLIC1 and CLIC2,
respectively.

\begin{figure}
\includegraphics[scale=0.7]{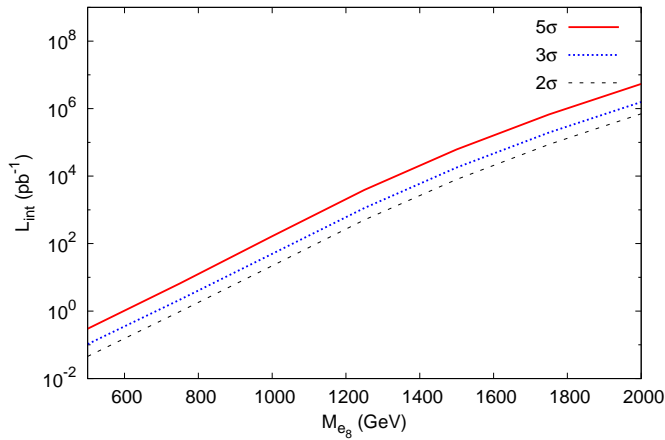}

\caption{The necessary integrated luminosity for the indirect observaion of
$e_{8}$ at the ILC.}

\end{figure}

\begin{figure}
\includegraphics[scale=0.7]{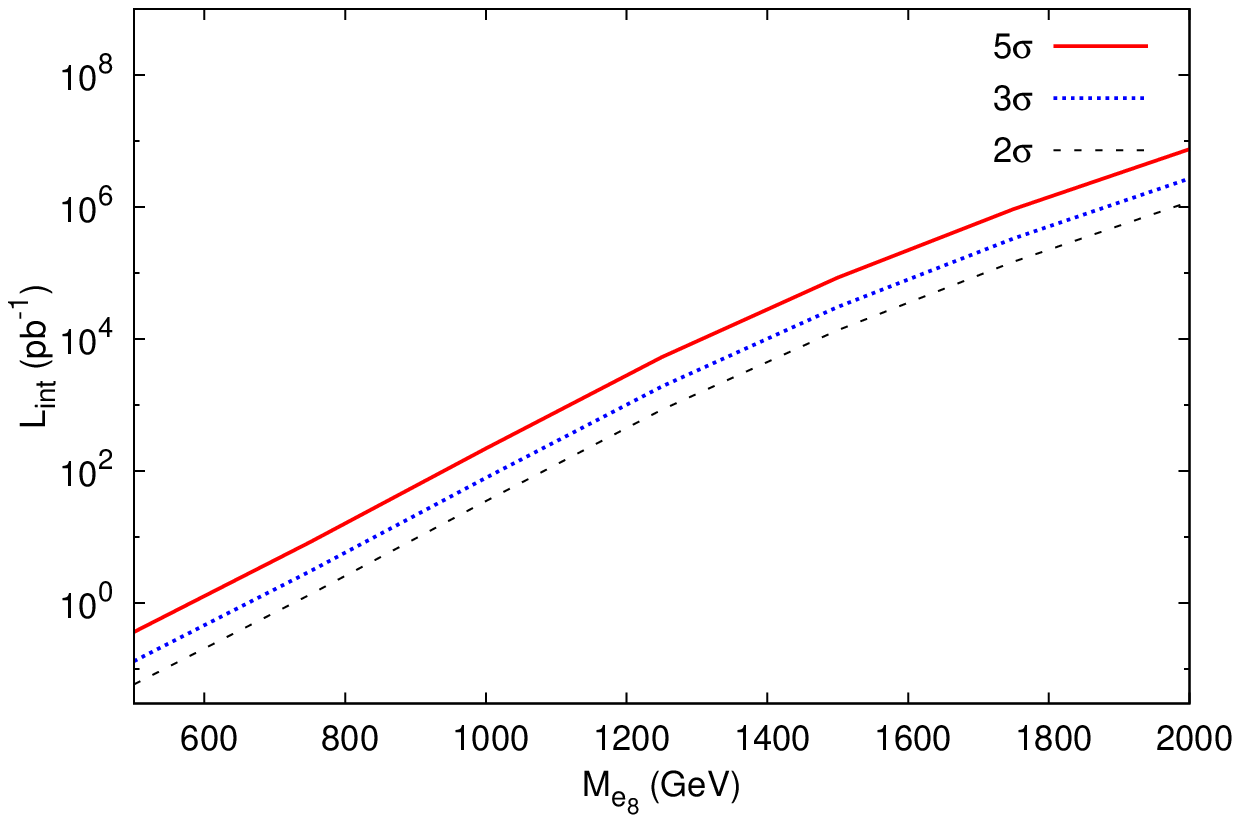}

\caption{The necessary integrated luminosity for the indirect observaion of
$e_{8}$ at the CLIC1. }

\end{figure}

\begin{figure}
\includegraphics[scale=0.7]{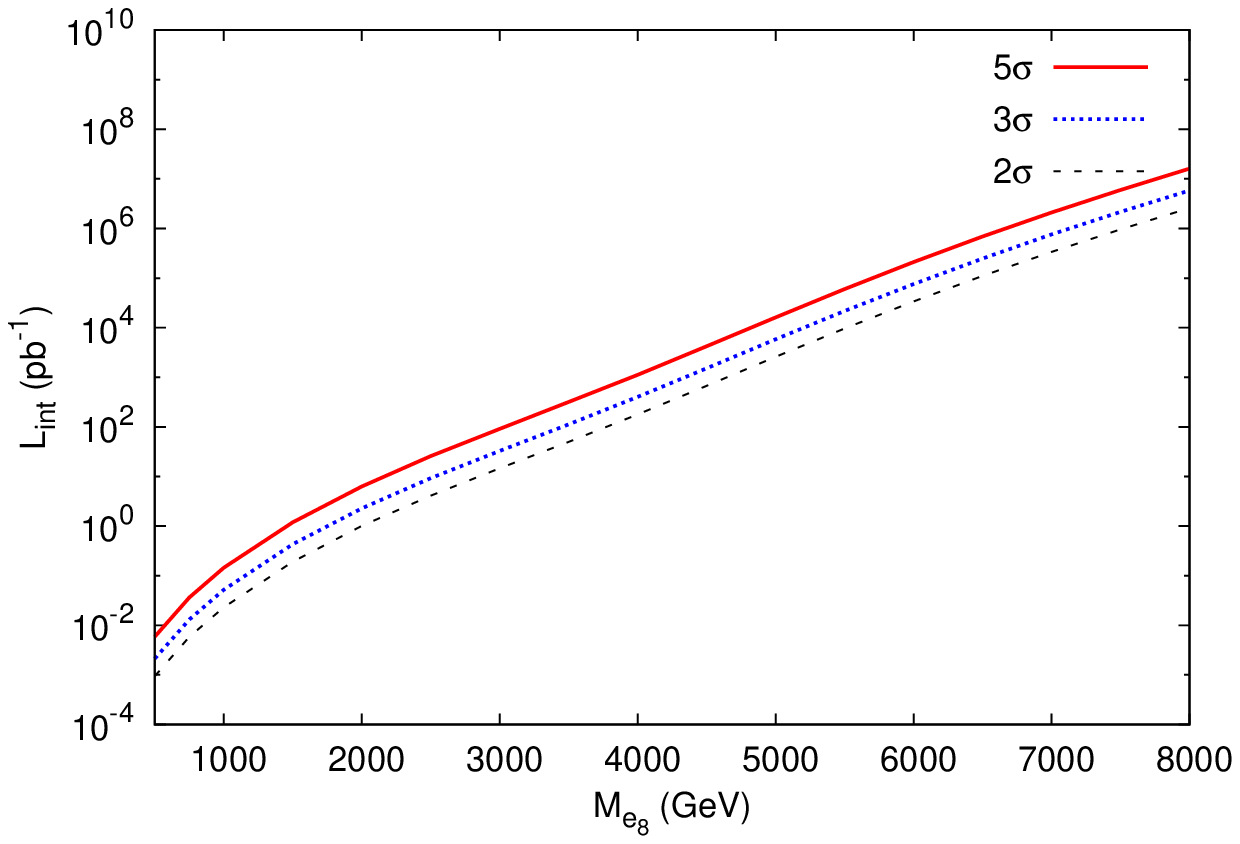}

\caption{The necessary integrated luminosity for the indirect observaion of
$e_{8}$ at the CLIC2.}

\end{figure}

The reachable $e_{8}$ mass values for one and three year operation
with nominal luminosities are given in Table IV.

\begin{table}
\begin{tabular}{|c|c|c|c|c|}
\hline 
Colliders  & years  & $5\sigma$  & $3\sigma$  & $2\sigma$\tabularnewline
\hline
\hline 
\multicolumn{1}{|c|||}{ILC} & $1$ year  & $1640$ GeV  & $1750$ GeV  & $1850$ GeV\tabularnewline
\hline 
ILC  & $3$ year  & $1760$ GeV  & $1880$ GeV  & $1980$ GeV\tabularnewline
\hline 
CLIC1  & $1$ year  & $1600$ GeV  & $1710$ GeV  & $1800$ GeV\tabularnewline
\hline 
CLIC1  & $3$ year  & $1720$ GeV  & $1840$ GeV  & $1920$ GeV\tabularnewline
\hline 
CLIC2  & $1$ year  & $6430$ GeV  & $6880$ GeV  & $7260$ GeV\tabularnewline
\hline 
CLIC2  & $3$ year  & $6930$ GeV  & $7400$ GeV  & $7810$ GeV\tabularnewline
\hline
\end{tabular}

\caption{Reachable $e_{8}$ mass values for indirect discovery, observation
and exclusion at the ILC, CLIC1 and CLIC2.}

\end{table}

Up to now, we assumed the compositeness scale equal to the mass of
color octet electron ($\Lambda=M_{e_{8}}$). In general case this
scale can be decoupled from the mass of new particles. For this reason
below we consider limits for the compositeness scale as a function
of the color octet electron mass. In this analysis we use $L_{int}=200,$
$230$ and $590$ $fb^{-1}$ for ILC, CLIC1 and CLIC2, respectively.
These values correspond to one year operation with nominal luminosities.

In Figure 9, 10 and 11 we plot reachable values of the compositeness
scale as a function of $M_{e_{8}}$ for ILC, CLIC1 and CLIC2, respectively.

\begin{figure}
\includegraphics[scale=0.7]{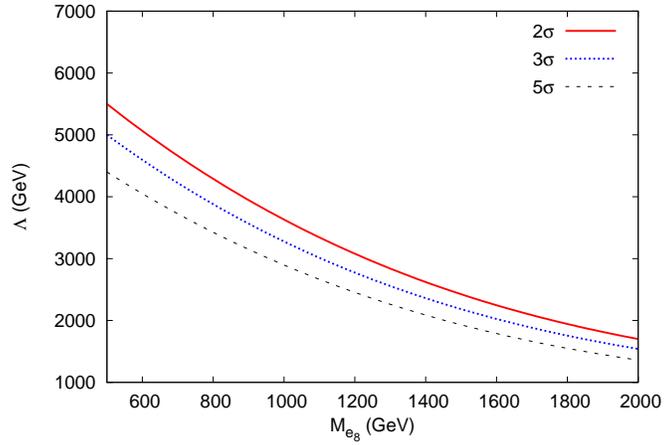}

\caption{Reachable values of the compositeness scale as a function of color
octet electron mass for ILC with $\sqrt{s}=0.5$ TeV and $L_{int}=200$
$fb^{-1}$.}

\end{figure}

\begin{figure}
\includegraphics[scale=0.7]{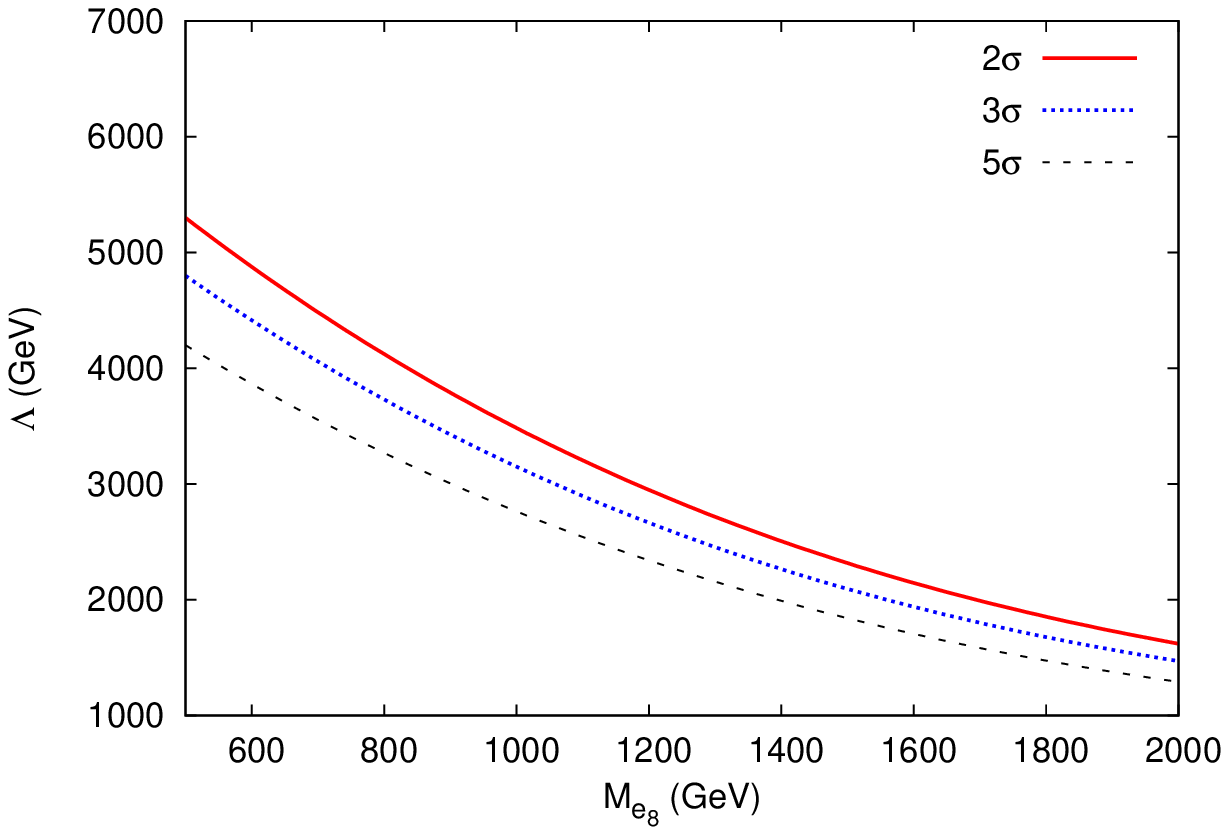}

\caption{Reachable values of the compositeness scale as a function of color
octet electron mass for CLIC1 with $\sqrt{s}=0.5$ TeV and $L_{int}=230$
$fb^{-1}$.}

\end{figure}

\begin{figure}
\includegraphics[scale=0.7]{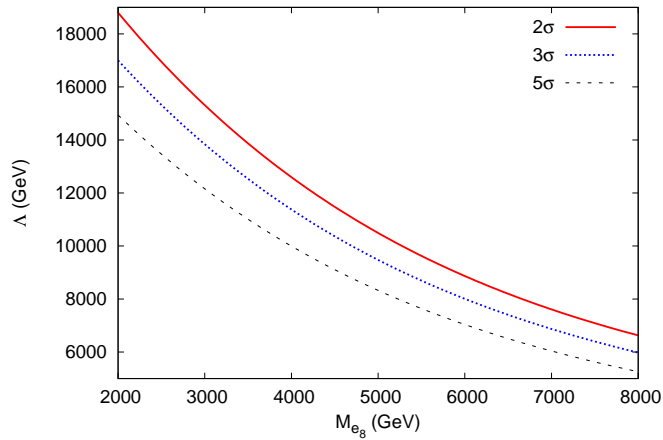}

\caption{Reachable values of the compositeness scale as a function of color
octet electron mass for CLIC2 with $\sqrt{s}=3$ TeV and $L_{int}=590$
$fb^{-1}$.}

\end{figure}

\section*{4. CONCLUSION}

We show that, if $\Lambda=M_{e_{8}}$ ILC will give opportunity for
indirect observation of octet electron upto $M_{e_{8}}=1750$ $(1880)$
GeV during one (three) operation with nominal luminosity. Corresponding
values for CLIC1 are about $40$ GeV lower. Therefore, both ILC and
CLIC1 will give opportunity to pass essentially the Tevatron capacity
for color octet electron ($e_{8}$) search. CLIC2 will be sensitive
to $4$ times higher mass values, namely, $e_{8}$ masses upto $6880$
($7400$) GeV could be observed during one (three) year operation
with nominal luminosity. These values essentially exceed capacity
of the LHC to observe color octet electron ($e_{8}$) via pair production.

If the compositeness scale and color octet electron mass are decoupled,
observable values of $\Lambda$ at ILC and CLIC1 exceeds $M_{e_{8}}$
for $M_{e_{8}}\lesssim1700$ GeV (at $M_{e_{8}}=500$ GeV the ratio
$\Lambda/M_{e_{8}}\approx10$). At the CLIC2 $\Lambda$ exceeds mass
for $M_{e_{8}}\lesssim7000$ GeV (at $M_{e_{8}}=2000$ GeV observable
$\Lambda$ is $17000$ GeV).

\section*{Acknowledgments}

This work is supported by TUBITAK in the framework of the BIDEP 2218
post-doctoral program.

\end{document}